\documentstyle[pra,aps,graphicx]{revtex}
\def\mathbf{\vec}

\def\ca{\c{c}\~{a}}

\begin{document}
\draft
\title{Path Integral Bosonization of the 't Hooft Determinant:
       Fluctuations and Multiple Vacua}
\author{Alexander A. Osipov\thanks{On leave from the Joint Institute for
 Nuclear Research, Laboratory of Nuclear Problems, 141980 Dubna,
        Moscow Region, Russia.} and Brigitte Hiller}
\address{Centro de F\'{\i}sica Te\'{o}rica, Departamento de
         F\'{\i}sica da Universidade de Coimbra, 3004-516 Coimbra, Portugal}
\date{\today}
\maketitle

\begin{abstract}
The 't Hooft six quark flavor mixing interaction ($N_f=3$) is bosonized by 
the path integral method. The considered complete Lagrangian is constructed 
on the basis of the combined 't Hooft and $U(3) \times U(3)$ extended
chiral four fermion Nambu -- Jona-Lasinio interactions. The method of
the steepest descents is used to derive the effective mesonic Lagrangian.
Additionally to the known lowest order stationary phase (SP) result of 
Reinhardt and Alkofer we obtain the contribution from the small quantum 
fluctuations of bosonic configurations around their stationary phase 
trajectories. It affects the vacuum state of hadrons at low energies: 
whereas without the inclusion of quantum fluctuations the vacuum is uniquely 
defined for a fixed set of the model parameters, fluctuations give rise to 
multivalued solutions of the gap equations, marked at instances by drastic 
changes in the quark condensates. We derive the new gap equations and
analyse them in comparison with known results. We classify the solutions 
according to the number of extrema they may accomodate. We find up to 
four solutions in the $0<m_u, m_s<3$ GeV region.
\end{abstract}

\section{Introduction}
The global $U_L(3)\times U_R(3)$ chiral symmetry of the QCD Lagrangian 
(for massless quarks) is broken by the $U_A(1)$ Adler-Bell-Jackiw 
anomaly of the $SU(3)$ singlet axial current $\bar{q}\gamma_\mu\gamma_5q$.
Through the study of instantons \cite{Hooft:1976,Diakonov:1995},
it has been realized that this anomaly has physical effects with the
result that the theory contains neither a conserved $U(1)$ quantum
number, nor an extra Goldstone boson. Instead the effective $2N_f$ quark
interactions arise, which are known as 't Hooft interactions. In the
case of two flavors they are four-fermion interactions,
and the resulting low-energy theory resembles the old Nambu --
Jona-Lasinio model \cite{Nambu:1961}. In the case of three flavors
they are six-fermion interactions which are responsible for the correct
description of $\eta$ and $\eta'$ physics, and additionally lead to the
OZI-violating effects \cite{Bernard:1988,Kunihiro:1988},
\begin{equation}
\label{Ldet}
    {\cal L}_{\mbox{det}}=\kappa (\mbox{det}\bar{q}P_Rq
                                 +\mbox{det}\bar{q}P_Lq)
\end{equation}
where the matrices $P_{R,L}=(1\pm\gamma_5)/2$ are projectors and determinant
is over flavor indices.  

The physical degrees of freedom of QCD at low-energies are mesons. The
bosonization of the effective quark interaction (\ref{Ldet}) by the
path integral approach has been considered in \cite{Reinhardt:1988}, where 
the lowest order stationary phase approximation (SPA) has been used to 
estimate the leading contribution from the 't Hooft determinant.
In this approximation the functional integral is dominated by the 
stationary trajectories $r_{\mbox{st}}(x)$, determined by the extremum 
condition $\delta S(r)=0$ of the action $S(r)$. The lowest order SPA  
corresponds to the case in which the integrals associated with 
$\delta^2 S(r)$, for the path $r_{\mbox{st}}(x)$ are neglected and
only $S(r_{\mbox{st}})$ contributes to the generating functional. 
The next natural step in this scenario is to complete the semiclassical 
result of Reinhardt and Alkofer by including the contribution from the 
integrals associated with the second functional derivative 
$\delta^2 S(r_{\mbox{st}})$, and this is the subject of our paper.

An alternative method of bosonizing the 't Hooft determinant has been 
reviewed in \cite{Diakonov:1998}. The special path integral
representation for the quark determinant has been obtained by considering
$N_c$ as an algebraically large parameter. One should not forget that 
the 't Hooft's determinant interactions are induced by instantons and only
can be written in the simple determinantal form (\ref{Ldet}) in the
limit of large number of colours -- otherwise the many-fermion 
interactions have a more complicated structure. For our calculations
it means, in particular, that the terms of the bosonized Lagrangian 
induced by the 't Hooft's determinant interaction (\ref{Ldet}) should
be of order $1$, corresponding to the standard rules of $N_c$
counting. In this respect it is worthwhile to note that we have found 
that the meson vertices, induced by the 't Hooft's determinant 
interactions in the lowest order SPA and the term from the integral
associated with the second functional derivative $\delta^2 S(r_{\mbox{st}})$ 
have the same $N_c$-order and should be considered on the same footing. 

\section{Path Integral Bosonization}
To be definite, let us consider the theory of the quark fields in four
dimensional Minkowski space, with dynamics defined by the Lagrangian
density
\begin{equation}
\label{totlag}
  {\cal L}={\cal L}_{\mbox{NJL}}+{\cal L}_{\mbox{det}}.
\end{equation}
The first term here is the extended version of the Nambu -- Jona-Lasinio 
(NJL) Lagrangian ${\cal L}_{\mbox{NJL}}={\cal L}_0+{\cal L}_{\mbox{int}}$, 
consisting of the free field part 
\begin{equation}
  {\cal L}_0=\bar{q}(i\gamma^\mu\partial_\mu -\hat{m})q,
\end{equation}
and the $U(3)_L\times U(3)_R$ chiral symmetric four-quark interaction  
\begin{equation}
  {\cal L}_{\mbox{int}}=\frac{G}{2}[(\bar{q}\lambda_aq)^2+
                        (\bar{q}i\gamma_5\lambda_aq)^2].
\end{equation}
We assume that quark fields have color and flavor indices running
through the set $i=1,2,3$; $\lambda_a$ are the standard $U(3)$ Gell-Mann 
matrices with $a=0,1,\ldots ,8$. The current quark mass, $\hat{m}$, is a 
nondegenerate diagonal matrix with elements $\mbox{diag}(\hat{m}_u, 
\hat{m}_d, \hat{m}_s)$, it explicitly breaks the global chiral 
$U(3)_L\times U(3)_R$ symmetry of the ${\cal L}_{\mbox{NJL}}$ Lagrangian. 
The second term in (\ref{totlag}) is given by (\ref{Ldet}). Letting  
\begin{equation}
\label{param}
   s_a=-\bar{q}\lambda_aq, \quad 
   p_a=\bar{q}i\gamma_5\lambda_aq, \quad 
   s=s_a\lambda_a, \quad
   p=p_a\lambda_a
\end{equation}
yields 
\begin{equation}
\label{mdet}
   {\cal L}_{\mbox{det}}=-\frac{\kappa}{64}\left[
   \mbox{det}(s+ip)+\mbox{det}(s-ip)\right]
\end{equation}
with determinants written in terms of the mesonic type quark bilinears.
This identity is a first step to the bosonization of the theory with
Lagrangian (\ref{totlag}). 

The dynamics of the system is described by the vacuum transition
amplitude in the form of the path integral
\begin{equation}
\label{genf1}
   Z=\int {\cal D}q{\cal D}\bar{q}\exp\left(i\int d^4x{\cal L}\right).
\end{equation}  
By means of a simple trick, suggested by Reinhardt and Alkofer, it is
easy to write down this amplitude as 
\begin{equation}
\label{genf2}
   Z=\int {\cal D}q{\cal D}\bar{q}{\cal D}\sigma_a
          {\cal D}\phi_a{\cal D}r_1^a{\cal D}r_2^a
     \exp\left(i\int d^4x{\cal L}'\right)
\end{equation}  
with
\begin{eqnarray}
\label{lagr1}
  {\cal L}'&=&\bar{q}(i\gamma^\mu\partial_\mu -\hat{m}-\sigma 
              +i\gamma_5\phi )q
              -\frac{1}{2G}\left[(\sigma_a)^2+(\phi_a)^2\right]
              +r_1^a(\sigma_a+G\bar{q}\lambda_aq)
              \nonumber \\
            &+&r_2^a(\phi_a-G\bar{q}i\gamma_5\lambda_aq)
              -\frac{\kappa}{(4G)^3}\left[
               \mbox{det}(\sigma+i\phi )
              +\mbox{det}(\sigma-i\phi )\right].
\end{eqnarray}
Eq.(\ref{genf2}) defines the same expression as Eq.(\ref{genf1}). To see
this one has to integrate first over auxiliary fields $r_1^a,\ r_2^a$. 
It leads to $\delta$-functionals which can be integrated out by taking
integrals over $\sigma_a$, and $\phi_a$, and which bring us back to the  
expression (\ref{genf1}). From the other side, it is easy to rewrite 
Eq.(\ref{genf2}) in a form appropriate to finish the bosonization, i.e., 
to calculate the integrals over quark fields and integrate out from $Z$ 
the unphysical part of the auxiliary $r_1^a,\ r_2^a$ scalar fields. Indeed, introducing new
variables $\sigma\rightarrow\sigma +Gr_1,\ \phi\rightarrow\phi +Gr_2$,
and after that $r_1\rightarrow 2r_1-\sigma /G ,\ r_2\rightarrow 
2r_2-\phi/G$ we have
\begin{equation}
\label{genf3}
   Z=\int {\cal D}\sigma_a{\cal D}\phi_a
          {\cal D}q{\cal D}\bar{q}
          \exp\left(i\int d^4x{\cal L}_q(\bar{q},q,\sigma ,\phi )\right)
     \int {\cal D}r_{1a}{\cal D}r_{2a}
          \exp\left(i\int d^4x{\cal L}_r(\sigma ,\phi ,r_1,r_2)\right)
\end{equation}  
where
\begin{eqnarray}
\label{lagr2}
  {\cal L}_q&=&\bar{q}(i\gamma^\mu\partial_\mu -\hat{m}-\sigma 
              +i\gamma_5\phi )q, \\
\label{lagr3}
  {\cal L}_r&=&2G\left[(r_{1a})^2+(r_{2a})^2\right]
              -2(r_{1a}\sigma_a+r_{2a}\phi_a)
              -\frac{\kappa}{8}\left[
               \mbox{det}(r_{1}+ir_{2})
              +\mbox{det}(r_{1}-ir_{2})\right].
\end{eqnarray}
The Fermi fields enter the action bilinearly, we can always integrate
over them, because in this case we deal with the standard Gaussian type 
integral. At this stage one should also shift the scalar fields 
$\sigma_a\rightarrow\sigma_a+\Delta_a$ by demanding that the vacuum 
expectation values of the shifted fields vanish $<0|\sigma_a|0>=0$.
In other words, all tadpole graphs in the end should sum to zero, giving 
us gap equations to fix parameters $\Delta_a$. Here $\Delta_a = m_a -
\hat{m}_a$, with $m_a$ denoting the constituent quark masses 
\footnote{The shift by the current quark mass is needed to hit the 
correct vacuum state, see e.g. \cite{Osipov:2001}.}. 
To evaluate path integrals 
over $r_{1,2}$ one has to use the method of stationary phase, or,
after the formal analytic continuation in the time coordinate $x_4=ix_0$,
the method of steepest descents. Let us consider this task in some detail. 

The Euclidean (imaginary time) version of the path integral under 
consideration is  
\begin{equation}
\label{intJ}
     J(\sigma ,\phi )=\int^{+\infty}_{-\infty}{\cal D}r_{1a}
     {\cal D}r_{2a}
     \exp\left(\int d^4x{\cal L}_r(\sigma ,\phi ,r_1,r_2)\right).
\end{equation} 
This integral is hopelessly divergent even if $\kappa=0$. One should
say at this point that we are not really interested in (\ref{intJ}) but
only in its analytic continuation. Let us suppose we analytically change 
${\cal L}_r(\sigma ,\phi ,r_1,r_2)$ in some way such that we go from this 
situation back to the one of interest. To keep the integral convergent, we 
must distort the contour of integration into the complex plane following 
the standard procedure of the method of the steepest descents. This method 
gives the first term in an asymptotic expansion of $J(\sigma ,\phi )$,
valid for $\hbar\rightarrow 0$. We lead the contour along the straight line 
which is parallel to the imaginary axis and crosses the real axis at the 
saddle point $r^a_{\mbox{st}}$. It is in the sense of this continuation    
that the integral $J(\sigma ,\phi )$ of (\ref{intJ}) is to be interpreted
as
\begin{equation}
\label{ancJ}
     J(\sigma ,\phi )=
     \int^{+i\infty+r_{\mbox{st}}}_{-i\infty +r_{\mbox{st}}}
     {\cal D}r_{1a}{\cal D}r_{2a}
     \exp\left(\int d^4x{\cal L}_r(\sigma ,\phi ,r_1,r_2)\right).
\end{equation} 
Near the saddle point $r^a_{\mbox{st}}$,
\begin{equation}
\label{lagr4}
  {\cal L}_r\approx {\cal L}_r(r_{\mbox{st}})
            +\frac{1}{2}\sum_{\alpha ,\beta }\tilde{r}_\alpha
            {\cal L}''_{\alpha\beta}(r_{\mbox{st}})\tilde{r}_\beta 
\end{equation}
where the saddle point, $r^a_{\mbox{st}}$, is a solution of the equations
${\cal L}'_r(r_1,r_2)=0$ determining a flat spot of the surface 
${\cal L}_r(r_1,r_2)$. 
\begin{equation}
\label{saddle}
  \left\{
         \begin{array}{rcl}
         2Gr^a_1-(\sigma +\Delta )_a
         -\frac{3\kappa}{8}A_{abc}(r_1^br_1^c-r_2^br_2^c)&=&0 \\
         2Gr^a_2-\phi_a+\frac{3\kappa}{4}A_{abc}r_1^br_2^c&=&0.
         \end{array}
  \right.
\end{equation}
This system is well-known from \cite{Reinhardt:1988}. The totally 
symmetric constants, $A_{abc}$, come from the definition of the 
flavor determinant: $\det r=A_{abc}r^ar^br^c$, and equal to
\begin{equation}
\label{A}
   A_{abc}=\frac{1}{3!}\epsilon_{ijk}\epsilon_{mnl}(\lambda_a)_{im}
           (\lambda_b)_{jn}(\lambda_c)_{kl}.
\end{equation}
They are closely related with the $U(3)$ constants $d_{abc}$. We use 
in (\ref{lagr4}) symbols $\tilde{r}^a$ for the differences
$(r^a-r^a_{\mbox{st}})$. To deal with the multitude of integrals in
(\ref{ancJ}) we define a column vector $\tilde{r}$ with eighteen
components $\tilde{r}_\alpha =(\tilde{r}^a_1, \tilde{r}^a_2)$ and
with the matrix ${\cal L}''_{\alpha\beta}(r_{\mbox{st}})$ being equal to
\begin{equation}
\label{Qab}
  {\cal L}''_{\alpha\beta}(r_{\mbox{st}})=4GQ_{\alpha\beta},
  \quad
  Q_{\alpha\beta}=\left(
  \begin{array}{cc}
  \delta_{ab}-\frac{3\kappa}{8G}A_{abc}r_{1\mbox{st}}^{c}
  &\frac{3\kappa}{8G}A_{abc}r_{2\mbox{st}}^{c}\\
  \frac{3\kappa}{8G}A_{abc}r_{2\mbox{st}}^{c}
  &\delta_{ab}+\frac{3\kappa}{8G}A_{abc}r_{1\mbox{st}}^{c}
  \end{array}
  \right).
\end{equation} 
Eq.(\ref{ancJ}) can now be concisely written as
\begin{equation}
\label{ancJ2}
     J(\sigma ,\phi )=\exp\left(\int d^4x {\cal L}_r(r_{\mbox{st}})
                      \right)
     \int^{+i\infty}_{-i\infty}
     {\cal D}\tilde{r}_{\alpha}
     \exp\left(2G\int d^4x\tilde{r}^{\mbox{t}}Q(r_{\mbox{st}})
     \tilde{r} 
     \right)\left[1+{\cal O}(\hbar )\right].
\end{equation} 

Our next task is to evaluate the integrals over $\tilde{r}_{\alpha}$.
Before we do this, though, some comments should be made about what we
have done so far: 

(1) The first exponential factor in Eq.(\ref{ancJ2}) is not new. 
It has been obtained by Reinhardt and Alkofer in \cite{Reinhardt:1988}. A 
bit of manipulation with expressions (\ref{lagr3}) and (\ref{saddle}) leads 
us to the result
\begin{equation}
\label{Lrst}
   {\cal L}_r(r_{\mbox{st}})=\frac{2}{3}\left\{
            G[(r_{1\mbox{st}}^a)^2+(r_{2\mbox{st}}^a)^2]
            -2[(\sigma +\Delta)_ar^a_{1\mbox{st}}
            +\phi_ar^a_{2\mbox{st}}]\right\}. 
\end{equation} 
One can try to solve Eqs.(\ref{saddle}) looking for solutions 
$r^a_{1\mbox{st}}$ and $r^a_{2\mbox{st}}$ in the form of increasing
powers in $\sigma_a , \phi_a$  
\begin{eqnarray}
\label{rst}
   r^a_{1\mbox{st}}&=&h_a+h_{ab}^{(1)}\sigma_b
       +h_{abc}^{(1)}\sigma_b\sigma_c
       +h_{abc}^{(2)}\phi_b\phi_c+\ldots \\
   r^a_{2\mbox{st}}&=&h_{ab}^{(2)}\phi_b
       +h_{abc}^{(3)}\phi_b\sigma_c
       +\ldots 
\end{eqnarray}
Putting these expansions in Eqs.(\ref{saddle}) one can obtain the series 
of selfconsistent equations to determine the constants $h_a$,
$h^{(1)}_{ab}$, and $h^{(2)}_{ab}$
\begin{eqnarray}
\label{ha}
   &&2Gh_a-\Delta_a-\frac{3\kappa}{8}A_{abc}h_bh_c=0, \\  
   &&2G\left(\delta_{ac}-\frac{3\kappa}{8G}A_{acb}h_b
     \right)h^{(1)}_{ce}=\delta_{ae}, \\
   &&2G\left(\delta_{ac}+\frac{3\kappa}{8G}A_{acb}h_b
     \right)h^{(2)}_{ce}=\delta_{ae}. 
\end{eqnarray}
The other constants can be obtained from these ones, for instance, we have
\begin{equation}
   h^{(1)}_{abc}=\frac{3\kappa}{8}h^{(1)}_{a\bar a}h^{(1)}_{b\bar b} 
                 h^{(1)}_{c\bar c}A_{\bar a\bar b\bar c}, \quad 
   h^{(2)}_{abc}=-\frac{3\kappa}{8}h^{(1)}_{a\bar a}h^{(2)}_{b\bar b} 
                 h^{(2)}_{c\bar c}A_{\bar a\bar b\bar c}, \quad 
   h^{(3)}_{abc}=-\frac{3\kappa}{4}h^{(2)}_{a\bar a}h^{(2)}_{b\bar b} 
                 h^{(1)}_{c\bar c}A_{\bar a\bar b\bar c}. 
\end{equation}   
As a result the effective Lagrangian (\ref{Lrst}) can be expanded in powers 
of meson fields. Such an expansion (up to the terms which are cubic in 
$\sigma_a, \phi_a$) looks like
\begin{equation}
\label{lam}
   {\cal L}_r(r_{\mbox{st}})=-2h_a\sigma_a
                             -h_{ab}^{(1)}\sigma_a\sigma_b  
                             -h_{ab}^{(2)}\phi_a\phi_b
                             +{\cal O}(\mbox{field}^3).
\end{equation}  

(2) Our result (\ref{ancJ2}) has been based on the assumption that
all eigenvalues of matrix $Q$ are positive. It is true, for instance,
if $\kappa =0$. It may happen, however, that some eigenvalues of $Q$ 
are negative for some range of parameters $G$ and $\kappa$. 
In these cases there are no conceptual difficulties, for from the very 
beginning we deal with well defined Gaussian integrals and the 
integration over the corresponding $\tilde{r}_\alpha$ simply does not require 
analytic continuation.

We now turn to the evaluation of the path integral in Eq.(\ref{ancJ2}). 
In order to define the measure ${\cal D}\tilde{r}_\alpha$ more 
accurately let us expand $\tilde{r}_\alpha$ in a Fourier series
\begin{equation}
   \tilde{r}_\alpha (x)=\sum^\infty_{n=1}c_{n,\alpha}\varphi_n(x),
\end{equation} 
assuming that suitable boundary conditions are imposed. The set of 
the real functions $\{\varphi_n(x)\}$ form an orthonormal and complete 
sequence 
\begin{equation}
\label{compl}
   \int d^4x\varphi_n(x)\varphi_m(x)=\delta_{nm},\quad
   \sum_{n=1}^\infty\varphi_n(x)\varphi_n(y)=\delta (x-y).
\end{equation}
Therefore
\begin{equation}
\label{ancJ3}
     \int{\cal D}\tilde{r}_{\alpha}
     \exp\left(2G\int d^4x\tilde{r}^{\mbox{t}}Q(r_{\mbox{st}})
     \tilde{r}\right)=\int dc_{n,\alpha}
     \exp\left\{2G\sum c_{n,\alpha}\lambda^{\alpha\beta}_{nm}c_{m,\beta}
     \right\}=\frac{C}{\sqrt{\det (2G\lambda^{\alpha\beta}_{nm})}}.
\end{equation}
The normalization constant $C$ is not important for the following. The
matrix $\lambda^{\alpha\beta}_{nm}$ is equal to
\begin{equation}
  \lambda^{\alpha\beta}_{nm}=\int d^4x\varphi_n(x)Q_{\alpha\beta}(x)
  \varphi_m(x).
\end{equation}
From (\ref{Qab}) and (\ref{compl}) it follows that
\begin{equation}
\label{Qab2}
   2G\lambda^{\alpha\beta}_{nm}
   =\left(
            \begin{array}{cc}
            h^{(1)-1}_{ac}&0\\
            0&h^{(2)-1}_{ac}
            \end{array}
    \right)_{\alpha\sigma}
            \left(\delta_{\sigma\beta}\delta_{nm}
            +\int d^4x\varphi_n(x)F_{\sigma\beta}(x)\varphi_m(x)
            \right)
\end{equation}
with
\begin{equation}        
            F_{\sigma\beta}=\frac{3\kappa}{4}A_{eba}\left(
                   \begin{array}{cc}
                   -h^{(1)}_{ce}(r^a_{1\mbox{st}}-h_a)&
                   h^{(1)}_{ce}r^a_{2\mbox{st}}\\
                   h^{(2)}_{ce}r^a_{2\mbox{st}}&
                   h^{(2)}_{ce}(r^a_{1\mbox{st}}-h_a)
                   \end{array}
                   \right)_{\sigma\beta}.  
\end{equation}
Only the matrix $F_{\sigma\beta}$ depends here on fields $\sigma , \phi$. 
By absorbing in $C$ the irrelevant field independent part of 
$2G\lambda^{\alpha\beta}_{nm}$, and expanding the logarithm in the
representation $\det (1+F)=\exp\mbox{tr}\ln (1+F)$, one can obtain 
finally for the integral in (\ref{ancJ2})
\begin{equation} 
\label{Sr}
  J(\sigma ,\phi )=C'e^{S_r}, \quad 
    S_r=\int d^4x\left\{   
      {\cal L}_r(r_{\mbox{st}})+
      \frac{1}{2}\sum_{n=1}^\infty\frac{(-1)^n}{n}\mbox{tr}\left[
      F^n_{\alpha\beta}(r_{\mbox{st}})\right]\sum_{m=1}^\infty
      \varphi_m(x)\varphi_m(x)\right\}.
\end{equation}
The sum over $m$ in this expression, however, is not well defined and 
needs to be regularized. One can regularize it by introducing a Gaussian 
cutoff $M$ damping the contributions from the large momenta $k^2$
\begin{equation}
   \sum_{m=1}^\infty\varphi_m(x)\varphi_m(x)=\delta (0)
   \sim
   \int^{\infty}_{-\infty}\frac{d^4k}{(2\pi )^4}
   \exp\left(-\frac{k^2}{M^2}\right) 
   =\frac{M^4}{16\pi^2}
\end{equation}  
This procedure does not decrease the predictability of the model, for
anyway one has to regularize the quark loop contributions in (\ref{genf3}).
Alternatively, following ideas presented in \cite{Jackiw:2000}, one
can introduce the Ansatz: 
\begin{equation} 
\label{Sra} 
    S_r=\int d^4x\left\{   
      {\cal L}_r(r_{\mbox{st}})+
      \frac{a}{2G^2}\sum_{n=1}^\infty\frac{(-1)^n}{n}\mbox{tr}\left[
      F^n_{\alpha\beta}(r_{\mbox{st}})\right]\right\}
\end{equation}
proposing that the undetermined dimensionless constant $a$ will be fixed
by confronting the model with experiment afterwards.

\section{The Ground State}
Let us study the ground state of the model under consideration, then 
properties of the excitations will follow naturally.
To make further progress let us note that Eqs.(\ref{ha}) have non-trivial
solutions for $h_0,h_3,h_8$, corresponding to the spontaneous breaking
of chiral symmetry in the physical vacuum state with order parameters 
$\Delta_i\neq 0\quad (i=u,d,s)$. We may then use this fact to rewrite
Eqs.(\ref{ha}) as a system of only three equations to give $h_i$
\begin{equation} 
\label{hi}
   2Gh_i-\Delta_i=\frac{\kappa}{8}t_{ijk}h_jh_k
\end{equation}
where the totally symmetric coefficients $t_{ijk}$ are equal to zero
except for $t_{uds}=1$. They are related to coefficients $A_{abc}$ by
the embedding formula $3\omega_{ia}A_{abc}e_{bj}e_{ck}=t_{ijk}$ where 
matrices $\omega_{ia}$, and $e_{ai}$ are defined as follows

\begin{figure}[h]
  \begin{center}
   \begin{tabular}{cc}
     \resizebox{8cm}{!}{\includegraphics{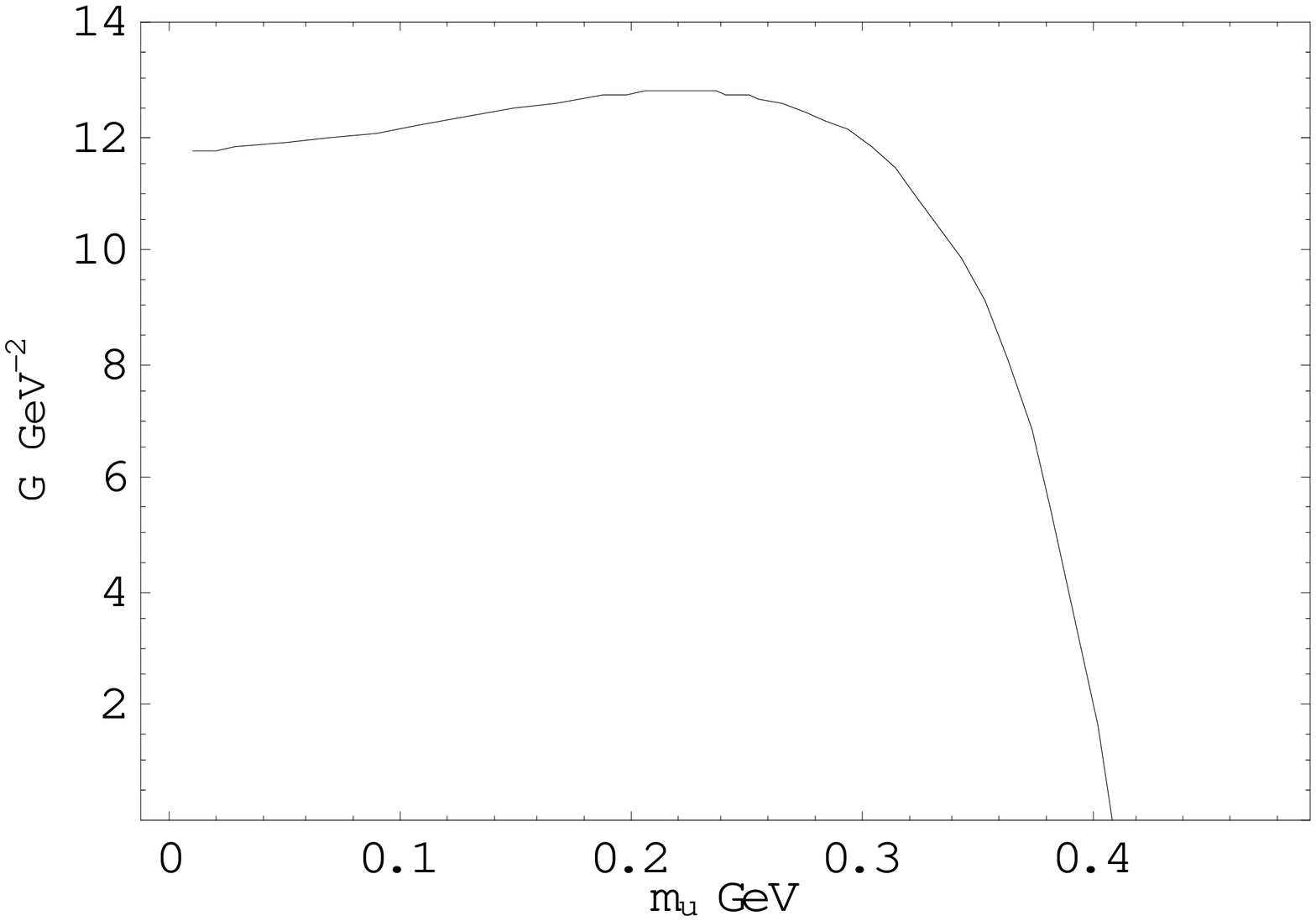}}
     &
     \resizebox{8.5cm}{!}{\includegraphics{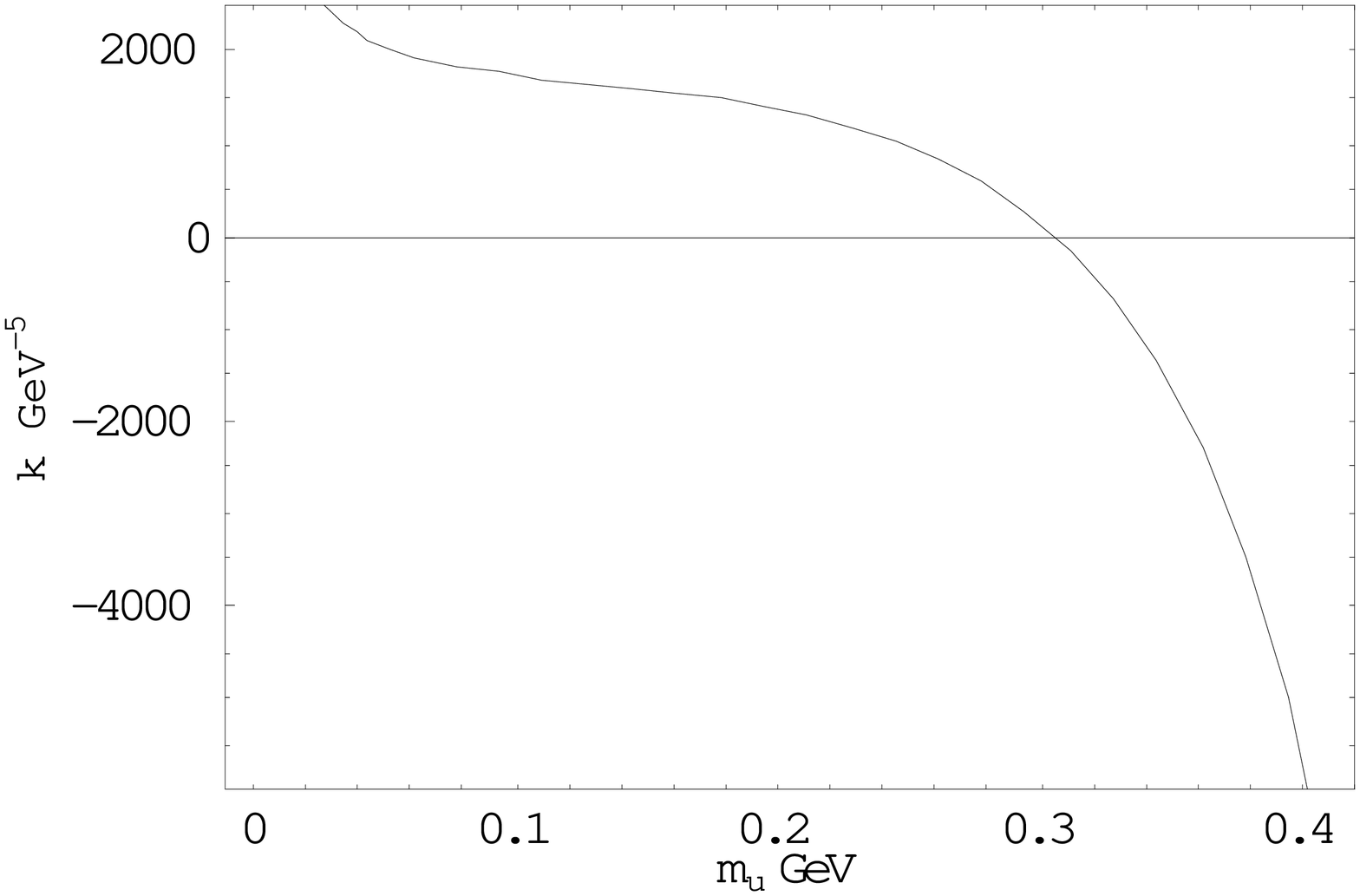}}\\
   \end{tabular}
  \end{center}
\caption{The couplings $G$ (left panel) and $\kappa$ (right panel) as 
functions of $m_u$ at fixed values of $m_s=572$ MeV and other
parameters in Eq.(\ref{Gk}). These curves show the typical $m_u$-dependence 
of the functions if one neglects the fluctuations term in Eq.(\ref{gap}).} 
\label{fig1}
\end{figure}

\begin{equation}
   e_{ai}=\frac{1}{2\sqrt 3}\left(
          \begin{array}{ccc}
          \sqrt 2&\sqrt 2&\sqrt 2 \\
          \sqrt 3&-\sqrt 3& 0 \\
          1&1&-2 
          \end{array} \right),\qquad
   \omega_{ia}=\frac{1}{\sqrt 3}\left(
          \begin{array}{ccc}
          \sqrt 2&\sqrt 3& 1 \\
          \sqrt 2&-\sqrt 3& 1 \\
          \sqrt 2&0&-2 
          \end{array} \right).
\end{equation}
Here the index $a$ runs $a=0,3,8$ (for the other values of $a$ the 
corresponding matrix elements are equal to zero). We have also
$h_a=e_{ai}h_i$, and $h_i=\omega_{ia}h_a$. Similar relations 
can be obtained for $\Delta_i$ and $\Delta_a$. In accordance with these 
notations we will use, for instance, that $h^{(1)}_{ci}=
\omega_{ia}h^{(1)}_{ca}$.

A tadpole graphs calculation gives for the gap equations the following 
result
\begin{equation}
\label{gap}
   2h_i+\frac{3a\kappa}{8G^2}
        \left(h^{(2)}_{ab}-h^{(1)}_{ab}\right)A_{abc}
        h^{(1)}_{ci}=\frac{N_c}{2\pi^2}m_iJ_0(m_i^2)
\end{equation}
where the left hand side is the  contribution from (\ref{Sra}) and the right
hand side is the contribution of the quark loop from (\ref{genf3})
with a regularized quadratically divergent integral $J_0(m^2)$ being
defined as
\begin{equation}
  J_0(m^2)=\int_0^\infty\frac{dt}{t^2}e^{-tm^2}\rho
           (t,\Lambda^2),\qquad
  \rho (t,\Lambda^2)=1-(1+t\Lambda^2)\exp (-t\Lambda^2).
\end{equation}

\begin{figure}[h]
  \begin{center}
   \begin{tabular}{cc}
     \resizebox{8.5cm}{!}{\includegraphics{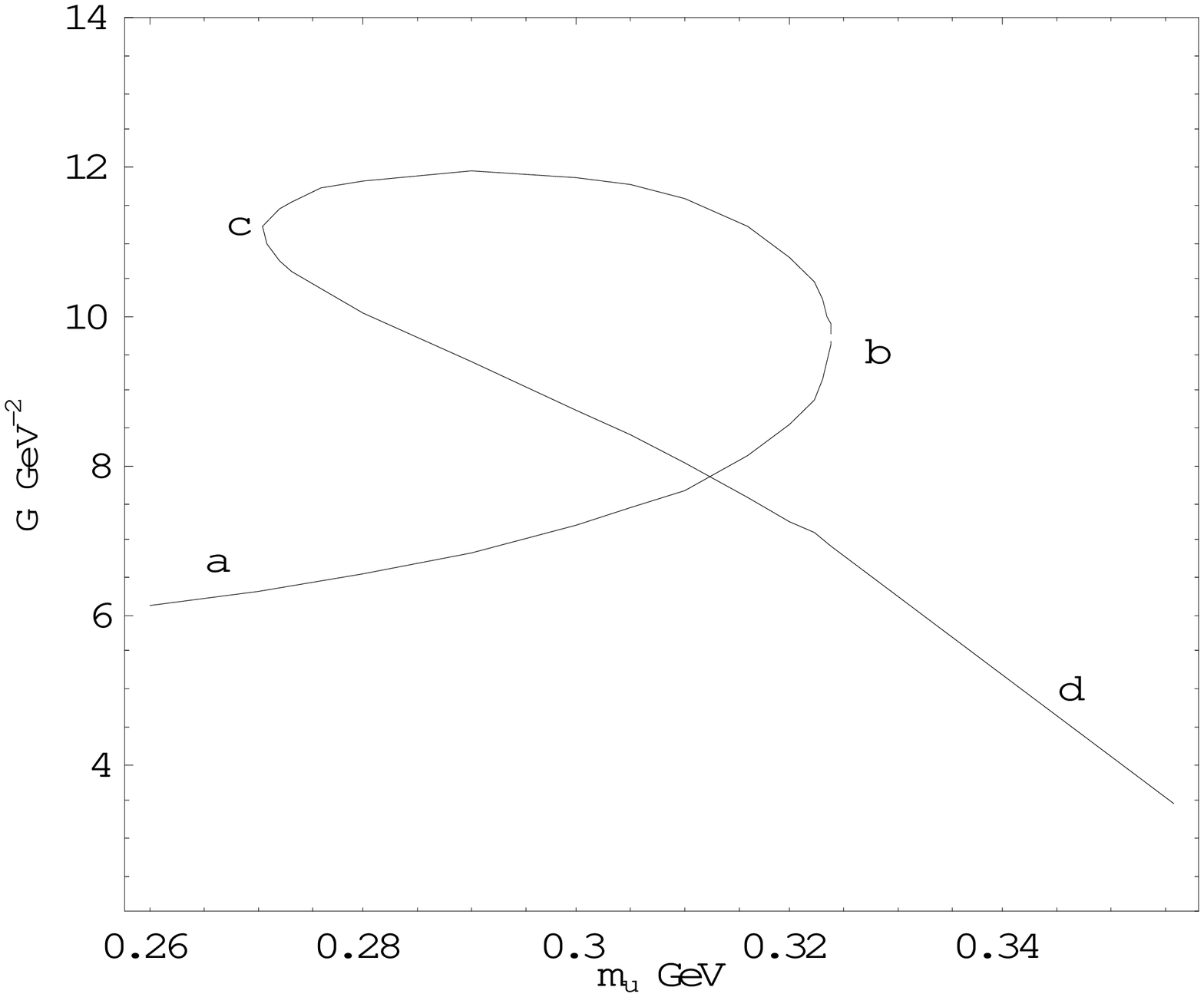}}
     &
     \resizebox{8.5cm}{!}{\includegraphics{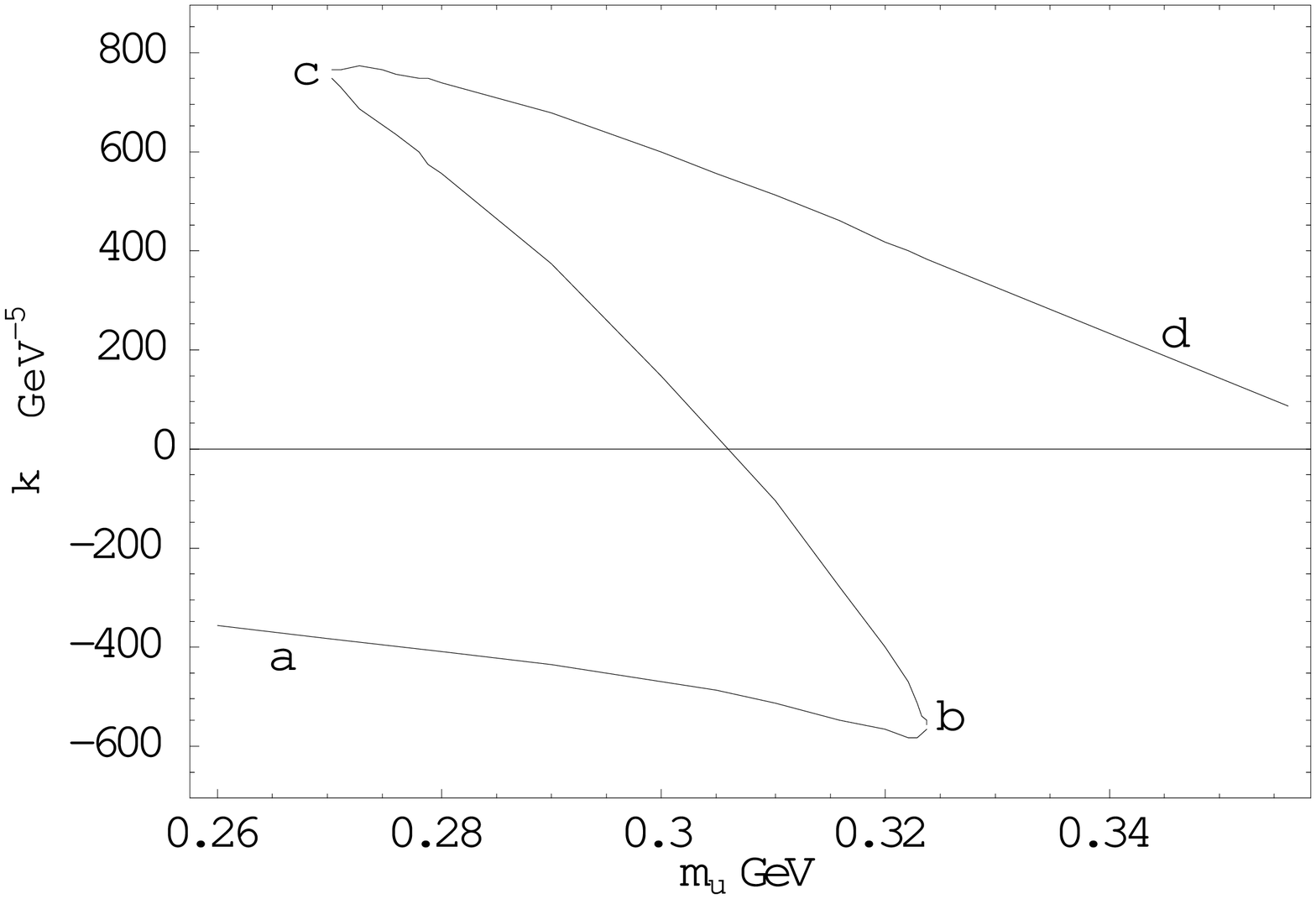}}\\
   \end{tabular}
  \end{center}
\caption{The couplings $G$ (left panel) and $\kappa$ (right panel) as 
functions of $m_u$ for the same (as in Fig.1) values of other fixed 
parameters and for the case when the fluctuations term in Eq.(\ref{gap})
is taken into account.} 
\label{fig2}
\end{figure}

\begin{figure}[h]
  \begin{center}
   \begin{tabular}{cc}
     \resizebox{8.5cm}{!}{\includegraphics{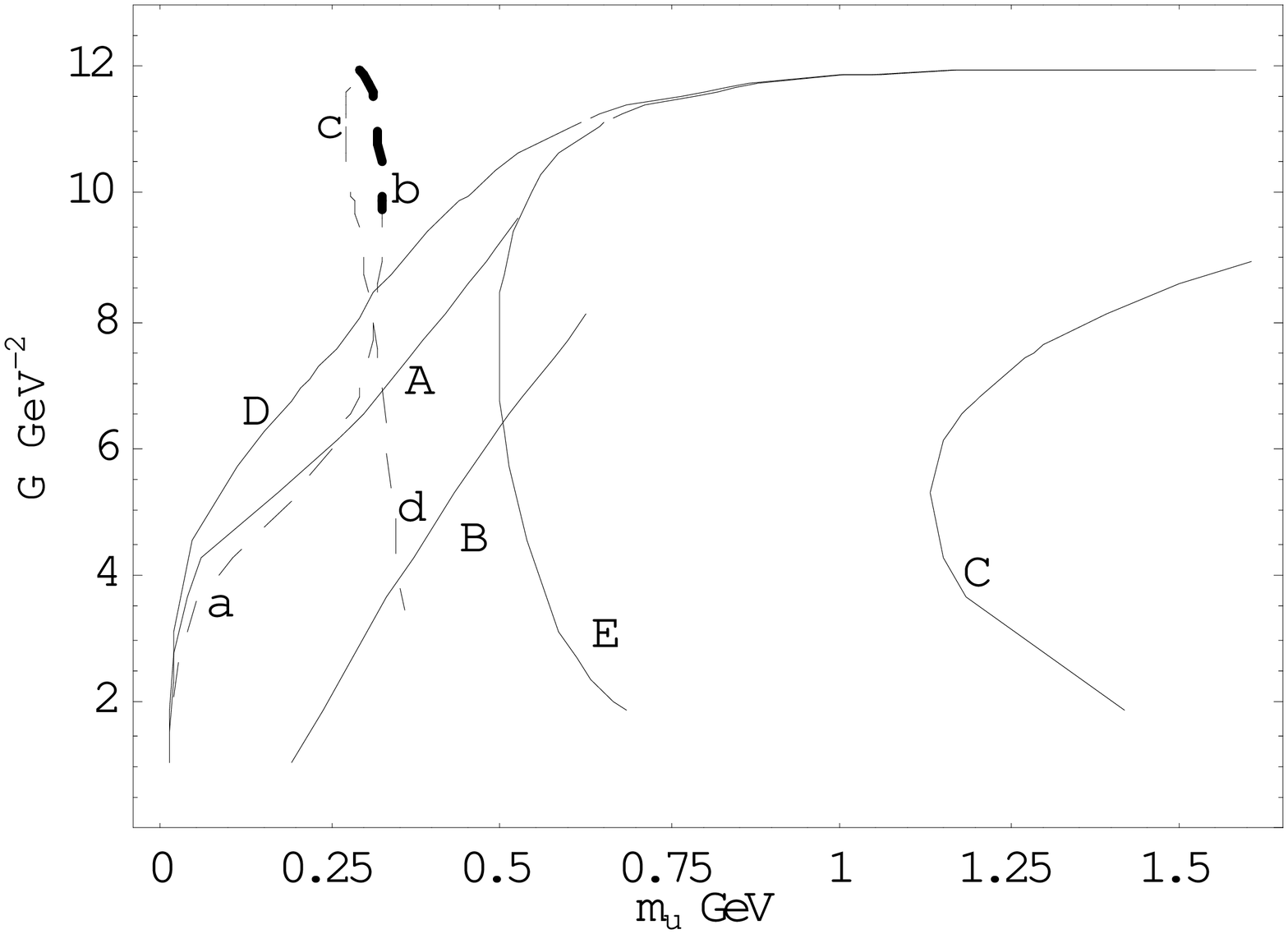}}
     &
     \resizebox{8.5cm}{!}{\includegraphics{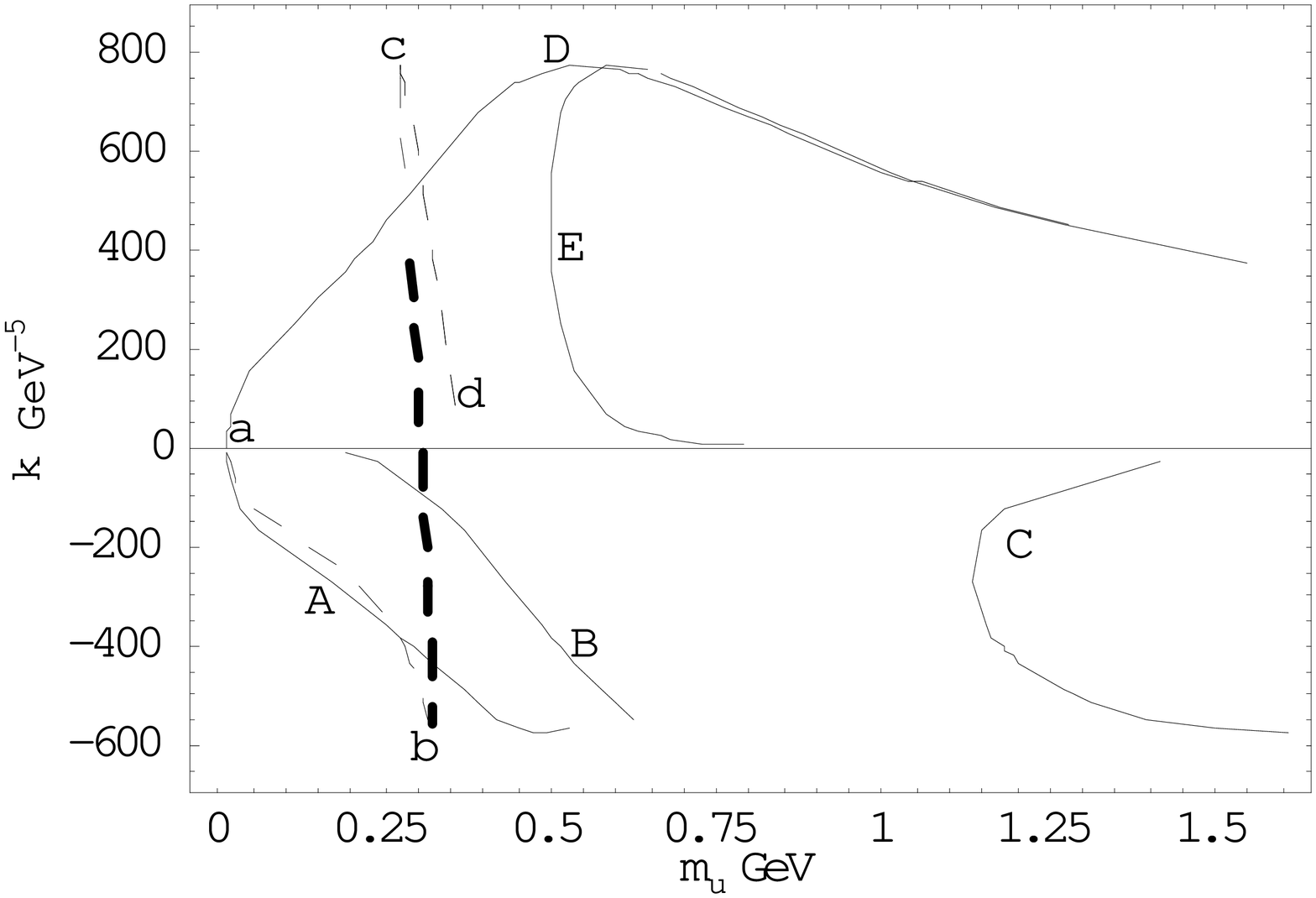}}\\
   \end{tabular}
  \end{center}
\caption{Different classes of $m_u$ solutions of the gap equations,
including fluctuations, at fixed $G, \kappa$ values. Four branches of
solutions, $A,B,C$ and $a-b$ stretch have negative $\kappa$. Three
branches, $D,E$ and along $c-d$ arm correspond to positive
$\kappa$. Bold dashes indicate that only one solution exists in this
region of $G, \kappa$. See further details in text.}
\label{fig3} 
\end{figure}

The second term on the left hand side of Eq.(\ref{gap}) is the 
correction resulting from the Gaussian integrals of the steepest
descent method, comprising the effects of small fluctuations around 
the stationary path. If one puts for a moment 
$a=0$ in Eq.(\ref{gap}), and combines the result with Eqs.(\ref{hi}), 
one finds gap equations which are very similar to the ones obtained in
\cite{Bernard:1988} (see equation (2.12) therein). At fixed input 
parameters  $G,\kappa ,\Lambda$ of the model, the gap equations can be
solved giving us the constituent quark masses $m_i$ as functions of the
current quark masses, $m_i=m_i(\hat{m}_j)$. Alternatively, by fixing
$\hat{m}_i$, one can obtain from the gap equations the non-trivial 
solutions $m_i=m_i(G,\kappa ,\Lambda )$. In particular, when 
$\hat{m}_u=\hat{m}_d$, these equations can be solved for $G$ and 
$\kappa$, giving expressions 
\begin{equation}
\label{Gk}
 G=\left(\frac{2\pi^2}{N_c}\right)
   \frac{m_u\Delta_uJ_0(m^2_u)-m_s\Delta_sJ_0(m^2_s)}{
         m_u^2J_0^2(m^2_u)-m_s^2J_0^2(m^2_s)},\qquad
 \kappa=\left(\frac{8\pi^2}{N_c}\right)^2
   \frac{m_s\Delta_uJ_0(m^2_s)-m_u\Delta_sJ_0(m^2_u)}{
   m_uJ_0(m_u^2)[      
   m_u^2J_0^2(m^2_u)-m_s^2J_0^2(m^2_s)]}.
\end{equation}
In Fig.\ref{fig1} we plot the curves of $G$ and $\kappa$ versus $m_u$ 
keeping constant $\Lambda =0.87\ \mbox{GeV},\ m_s=572\ \mbox{MeV},\ 
\hat{m}_s=200\ \mbox{MeV},\ \hat{m}_u=6\ \mbox{MeV}$. One can readily 
see that at given values for $(m_u,m_s)$ the curves yield 
unique values of $(G,\kappa )$, i.e. the vacuum state is well-defined
in this case.    

\begin{figure}[h]
  \begin{center}
   \begin{tabular}{cc}
     \resizebox{8.5cm}{!}{\includegraphics{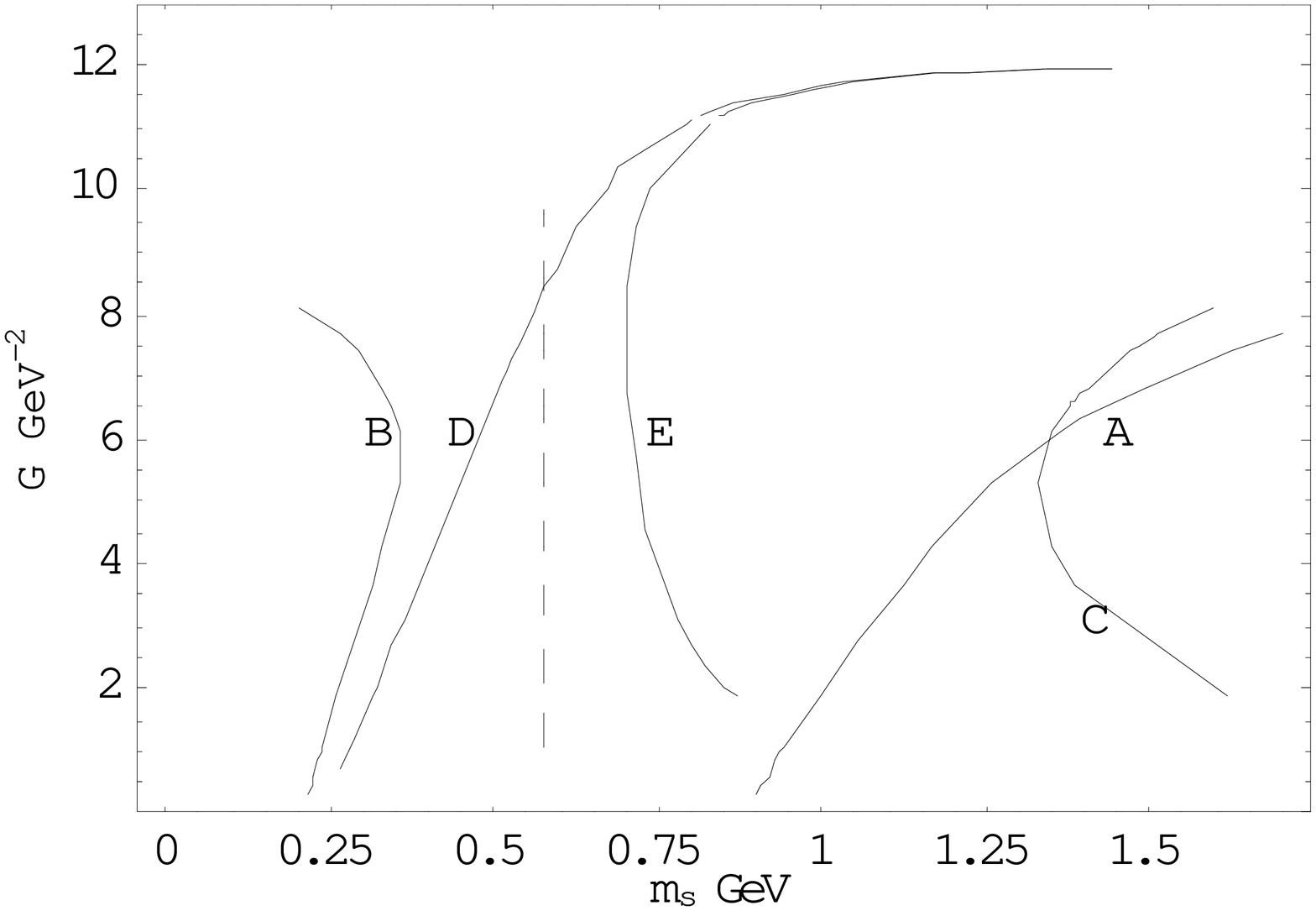}}
     &
     \resizebox{8.5cm}{!}{\includegraphics{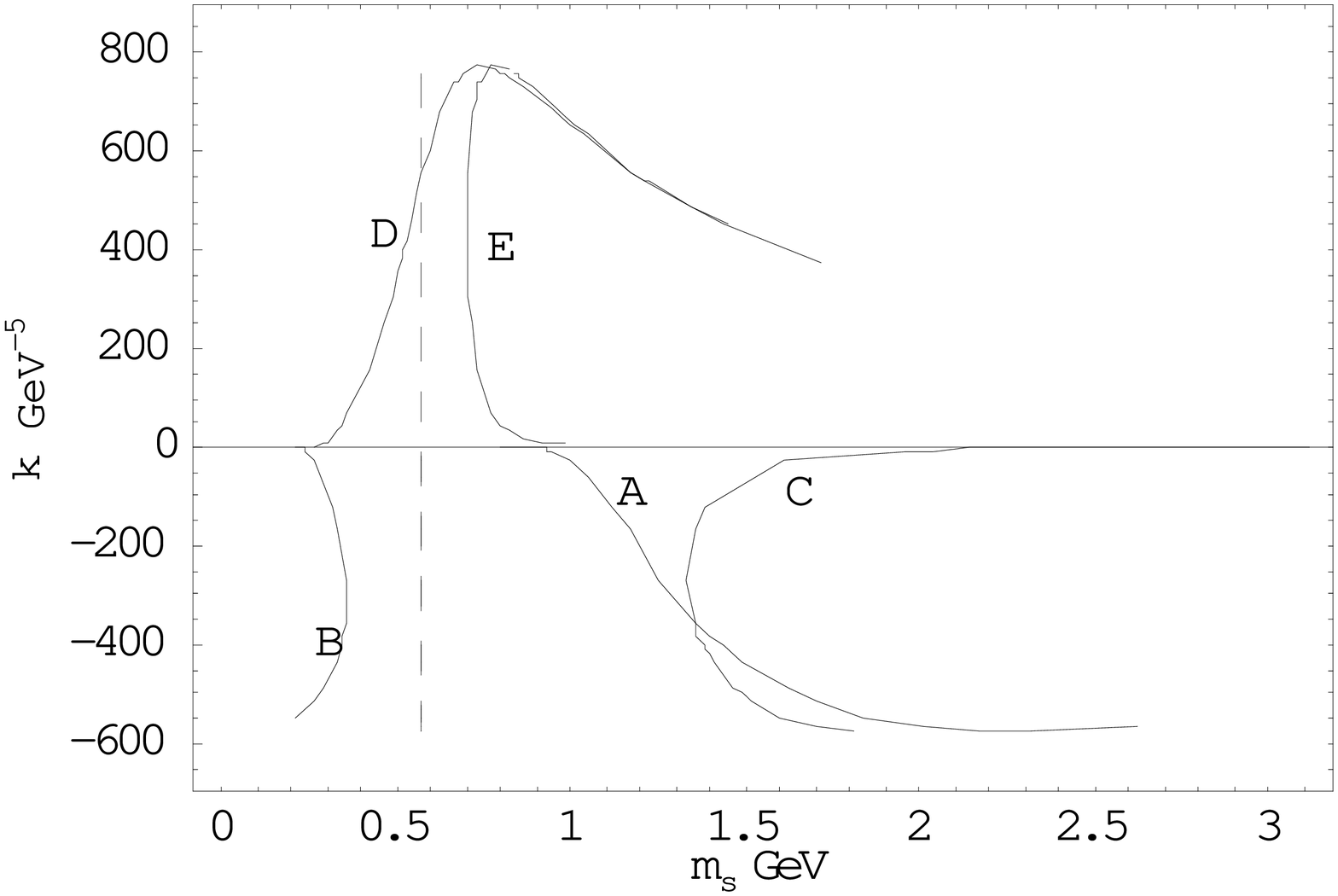}}\\
   \end{tabular}
  \end{center}
\caption{The same as in figure 3 for the $m_s$ solutions. The dashes
are the constant $m_s$ corresponding to the $abcd$ curve of 
Figs. 2 and 3.}\label{fig4} 
\end{figure}

Let us consider now the general case which we have when $a>0$ in 
Eq.(\ref{gap}). To illustrate the qualitative difference with the previous
case we put for definiteness $a=1$ and look again for the solutions 
$G=G(m_u)$ and $\kappa =\kappa (m_u)$ with the same set of fixed parameters.
The corresponding curves are plotted in Fig.\ref{fig2}. If the mass $m_u$ is 
sufficiently low that $m_u<m_u^{(min)}$, (in the figures denoted by
the region left the turning point $c$) or sufficiently high that
$m_u^{(max)}<m_u<m_s$, (in the figures denoted by the region right to the 
turning point $b$), then there exists again a single solution
with unique values of $(G,\kappa )$. However, there is now a 
region for $m_u$ in which $m_u^{(min)}<m_u<m_u^{(max)}$, where three 
values of couplings $(G,\kappa)$ are possible. 

Conversely, one can study the solutions: $m_u=m_u(G,\kappa ),\ 
m_s=m_s(G,\kappa$ at fixed values of input parameters: 
$\Lambda ,\hat{m}_u=\hat{m}_d,\hat{m}_s)$. As starting input values for 
the couplings $G$ and $\kappa$ in the gap equations we take the ones 
already determined along the path $abcd$  shown in Fig.\ref{fig2}, obtained 
at a constant value of the strange quark mass, $m_s=572$ MeV. For a chosen 
value of the set $(G,\kappa )$ we then search for further solutions 
$(m_u,m_s)$ of the Eq.(\ref{gap}), displaying results in Fig.\ref{fig3} 
for $m_u$ and in Fig.\ref{fig4} for $m_s$, correspondingly.  
The dashed curves are the repetition of the solutions encountered in 
Fig.\ref{fig2}. The bold dashes in Fig.\ref{fig3} indicate that we only 
find this one solution at fixed values of $G,\kappa$. Combining the 
information of Figs.\ref{fig3} and \ref{fig4}, one sees that one has
up to four solutions at fixed $G,\kappa$. Indeed, travelling along the 
original path $abcd$ one observes the following: the branch $ab$ is 
accompanied by three other branches, marked as $A,B$ and $C$ which
belong to the same class of solutions. One sees that these solutions
have negative $\kappa$ values.  From the turning point $b$ until the
maximum value of $G$ (and corresponding $\kappa$) (bold dashes) we
have no other solutions to the gap equations, as already stated. From 
this maximum $G$ value to the turning point $c$ and further along the 
$cd$ arm we encounter further two branches, denoted by $D$ and $E$ to 
the solutions of Eqs.(\ref{gap}) with same $G,\kappa$. They are positive 
$\kappa$ solutions. 

This very rich structure of the vacuum solutions implies the 
possibility of having several different values of the quark
condensates for the same $G,\kappa$ parameters, embracing also the
possibility which has been considered in connection with generalized
chiral perturbation theory (see e.g. section 4 in
\cite{Leutwyler:1996}).
We give here only a few examples. At $G=4.54\ \mbox{GeV}^{-2}, \kappa=153.04\
\mbox{GeV}^{-5}$ we have three solutions. The solution $m_u=346\ \mbox{MeV}, 
m_s=572\ \mbox{MeV}$, on the $cd$ arm has the quark condensates 
$<\bar{u}u>^{1/3}=-236.8\ \mbox{MeV}$,  $<\bar{s}s>^{1/3}=-183.5\ \mbox{MeV}$ 
and the ratio $R={(<\bar{s}s>/<\bar{u}u>)}^{1/3}=0.775$; the second solution, 
on the $E$ branch with $m_u=535\ \mbox{MeV}, m_s=732\ \mbox{MeV}$ has 
condensates $<\bar{u}u>^{1/3}=-249\ \mbox{MeV}$,  
$<\bar{s}s>^{1/3}=-183\ \mbox{MeV}$ and the ratio $R=0.738$; the third 
solution, located at the $D$ branch with $m_u=46\ \mbox{MeV},
m_s=418\ \mbox{MeV}$ has  condensates $<\bar{u}u>^{1/3}=-131\ \mbox{MeV}$,  
$<\bar{s}s>^{1/3}=-172\ \mbox{MeV}$ and $R=1.313$. We chose the next
example at $G=8.126\ \mbox{GeV}^{-2}, \kappa=-544.81\ \mbox{GeV}^{-5}$,
where there are four solutions. The solution $m_u=316\ \mbox{MeV},
m_s=572\ \mbox{MeV}$, on the $ab$ arm has the quark condensates 
$<\bar{u}u>^{1/3}=-233\ \mbox{MeV}$,  
$<\bar{s}s>^{1/3}=-184\ \mbox{MeV}$ and $R=0.787$; the second
solution, on the $B$ branch with $m_u=624\ \mbox{MeV}, m_s=205\
\mbox{MeV}$ has condensates $<\bar{u}u>^{1/3}=-249\ \mbox{MeV}$,  
$<\bar{s}s>^{1/3}=-56.7\ \mbox{MeV}$ and $R=0.227$; the third solution, 
located at the $A$ branch with $m_u=421\ \mbox{MeV}, m_s=1.84\ \mbox{GeV}$ 
has  condensates $<\bar{u}u>^{1/3}=-14.4\ \mbox{MeV}$,  
$<\bar{s}s>^{1/3}=-86\ \mbox{MeV}$ and $R=0.353$. The fourth solution, 
on the $C$ branch, with $m_u=1.394\ \mbox{GeV}, m_s=1.594\ \mbox{GeV}$
has condensates $<\bar{u}u>^{1/3}=-230\ \mbox{MeV}$,  
$<\bar{s}s>^{1/3}=-120\ \mbox{MeV}$ and the ratio $R=0.524$. As a final 
example we take the solutions at $G=11.96\ \mbox{GeV}^{-2}\simeq
G_{max}, \kappa=371.491\ \mbox{GeV}^{-5}$, where the branches $D$ and $E$ 
emerge and are very close to each other with $m_u=1.56\ \mbox{GeV}, 
m_s=1.72\ \mbox{GeV}$. The corresponding condensates are 
$<\bar{u}u>^{1/3}=-224\ \mbox{MeV}$,  
$<\bar{s}s>^{1/3}=-105\ \mbox{MeV}$ and  $R=0.467$. The other solution is
at the path $bc$ with $m_u=290\ \mbox{MeV}, m_s=572\ \mbox{MeV}$ with  
condensates $<\bar{u}u>^{1/3}=-229\ \mbox{MeV}$,  
$<\bar{s}s>^{1/3}=-184\ \mbox{MeV}$ and $R=0.8$.  

To summarize, we have found that in the presence of the 't Hooft 
interaction, treated beyound the lowest order SPA, several solutions to the
gap equations are possible at some range of input parameters, i.e. 
the same values of $G,\kappa ,\Lambda ,\hat{m}_i$ lead to different sets 
of constituent quark masses $(m_u, m_s)$ and, therefore, 
to different values of the quark condensates. A quite different
scenario emerges for the hadronic vacuum, which can now be multivalued. 
It makes our result essentially different from the ones obtained in 
\cite{Reinhardt:1988,Bernard:1988}. These findings must be further 
analysed in order to establish which of the extrema correspond to 
minima or maxima of the effective potential. This step will be done 
elsewhere in conjunction with the determination of the meson mass 
spectrum, as it also requires dealing with the terms with two powers 
of the meson fields in the ansatz of solutions Eq.(\ref{rst}) and in 
the related Lagrangian (\ref{lam}).

\section{Concluding remarks}
The purpose of this work has been twofold. Firstly we have developed
the technique which is necessary to go beyound the lowest order SPA
in the problem of the path integral bosonization of the 't
Hooft six quark interaction. We have shown how the pre-exponential
factor, connected with the steepest descent approach and which is
responsible for the quantum fluctuations around the
classical path, can be treated exactly, order by order, in a scheme of 
increasing number of mesonic fields, while preserving all chiral
symmetry requirements. This technique is rather general and can be
readily used in other applications. Second, we have explored with
considerable detail the implications of taking the quantum
fluctuations in account in the description of the hadronic vacuum. A
very complex multivalued vacuum emerges at fixed values of the input 
parameters $G$, $\kappa$, $\Lambda$ and current quark masses. We
encountered several classes of solutions. Searching in an interval of 
constituent quark masses from zero to $\simeq 3$GeV, we found
$G,\kappa$, regions caracterized by one, three and four solutions. 
The multiple vacua may have very interesting physical consequences and
applications.

\section*{Acknowledgements}
We are grateful to Dmitri Diakonov for valuable correspondence. We 
thank Dmitri Osipov and Pedro Costa for their help in converting the 
``Mathematica'' generated figures into the final ones. 
This work is supported by grants provided by Funda\ca o para a Ci\^encia e a
Tecnologia, POCTI/35304/FIS/2000 and NATO
"Outreach" Cooperation Program.

\baselineskip 12pt plus 2pt minus 2pt

\end{document}